\documentclass[journal,10pt,final]{IEEEtran}

\usepackage[T1]{fontenc}
\usepackage{cite}
\usepackage{url}
\usepackage{amsmath}
\usepackage{amssymb}
\usepackage[cmintegrals]{newtxmath}
\usepackage[labelformat=simple]{subcaption}

\usepackage{dblfloatfix}
\usepackage{graphicx}
\usepackage{epstopdf}
\usepackage{xcolor}
\usepackage{mathtools,amssymb,lipsum}
\usepackage[detect-weight=true, binary-units=true]{siunitx}
\usepackage[inline]{enumitem}
\usepackage{cuted}
\usepackage[figure]{algorithm2e} 
\usepackage{booktabs}

\usepackage{float}

\begin{document}

\title{A Markov Model of  Slice Admission Control}

\author{Bin~Han,~\IEEEmembership{Member,~IEEE,}
	    Di~Feng,
        and~Hans~D.~Schotten,~\IEEEmembership{Member,~IEEE}
\thanks{\textit{Bin Han and Hans D. Schotten are with University of Kaiserslautern.}}
	\thanks{\textit{Emails: \{binhan,schotten\}@eit.uni-kl.de}}%
	\thanks{\textit{D. Feng is with Universitat Aut\`onoma de Barcelona.}}%
	\thanks{\textit{Email: Di.Feng@e-campus.uab.cat}}%
}


\maketitle

\begin{abstract}
The emerging feature of network slicing in future Fifth Generation (5G) networks calls for efficient slice management. Recent studies have been focusing on the mechanism of slice admission control, which functions in a manner of state machine. This paper proposes a general state model for synchronous slice admission control, and proves it to be Markovian under a set of weak constraints. An analytical approximation of the state transition matrix to reduce computational complexity in practical applications is also proposed and evaluated.
\end{abstract}


\IEEEpeerreviewmaketitle

\setlength{\abovedisplayskip}{6pt}
\setlength{\belowdisplayskip}{6pt}

\section{Introduction}
\IEEEPARstart{N}{etwork} slicing\cite{alliance20155g} has been considered as an essential feature and one of the most important enablers of the Fifth Generation (5G) cellular communications networks. It allows mobile network operators (MNOs) to manage and utilize their physical and virtual network resources, i.e. the network infrastructure and the capacity of virtualized network functions (VNFs), in the form of logically independent virtual mobile networks, a.k.a. network slices. It provides broad improvements of scalability, flexibility, accountability, shareability and profitability to cellular networks\cite{rost2016mobile,rost2017network}.

One emerging challenge brought by network slicing is to efficiently allocate network resources over different slices towards better utility efficiency. More specifically, there are two typical scenarios of such inter-slice resource management. First, when an MNO directly provides services to end users and maintains these services on its own scalable slices, as proposed in \cite{han2017modeling}. Second, in the resource-sharing use case of so-called \emph{Slice as a Service} (SlaaS), which is discussed in \cite{sciancalepore2017slice}, an MNO packs its resources into standardized atomic slices and rents them to external tenants such as virtual MNOs (VMNOs) and service providers for agreement-based periodical revenue. In both scenarios, the MNO aims to maximize the overall network utility rate (e.g. the revenue rate) by adjusting its resource allocation subject to the constraints of resource pool limit and regulation rules.

Compared to the case of MNO's own service optimization, the SlaaS problem is more challenging due to the stochastic nature and fluctuating behavior of tenant demand for resources. Recently, multiple studies have been carried out on this topic \cite{bega2017optimising,caballero2017network,han2018slice}, applying the similar framework where a binary decision is made by the MNO for every slice admission, i.e. to accept or decline the tenant request for a new slice. Most of these work consider the system as a state machine, where state transitions are triggered by the MNO's responses to randomly arriving tenant requests. 

Despite the advantage of simple structure, such state-machine models suffer from two drawbacks. First, tenant requests by nature arrive in an asynchronous manner, leading to an asynchronous state model, while MNOs usually have synchronous frameworks for network controlling and management. Second, the models reported in literature only support simulative or exploitation-based evaluation of MNO's decision strategies. No analytical method of evaluation has been proposed. 

In this paper, we focus on an unstudied feature of such state models of slice admission systems: when operated in a synchronous method, i.e. when the MNO makes decision to tenant requests periodically instead of immediately upon request arrivals, is the derived synchronous state-model Markovian? Answering this question will help us to
\begin{enumerate*}
	\item better understand the system behavior towards efficient deployments of advanced techniques to optimize the MNO's decisions; and
	\item simplify the analytical evaluation of MNO's decision strategies.
\end{enumerate*}

The remainder of this paper is structured as follows: In Sec.~\ref{sec:system_model} we formulate the  model of asynchronous slice admission systems, defining critical concepts such as resource feasibility and slicing strategy. Then in Sec.~\ref{sec:sync_model} we consider the synchronous slice admission model, and prove it to be Markovian when the statistics of tenant requests are memoryless. Afterwards we provide the approximate analytical calculations of the state transition probabilities in both single step and long term as well. Sec.~\ref{sec:verification} numerically evaluates the feasibility of this Markov model and the accuracy of proposed calculation of the state transition matrix. To the end we close this paper with our conclusion and some discussions in Sec.~\ref{sec:conclusion}.

\section{System Model}\label{sec:system_model}
	\let\thefootnote\relax\footnotetext{\textit{This work was supported in part by the European Union Horizon-2020 Project 5G-MoNArch under Grant Agreement 761445 and in part by the Network for the Promotion of Young Scientists, University of Kaiserslautern (individual funding).}	\\{\copyright2018 IEEE}}
	\subsection{Resource Allocation and Resource Feasibility}\label{subsec:resource_model}
	A resource pool with $M$ different types of countable resources can be described with a vector $\mathbf{r}=[r_1,r_2,\dots,r_M]^\text{T}$. Consider $N$ different slice types, for every slice type $n\in[1,2,\dots,N]$ it costs a resource bundle $\mathbf{c}_n=[c_{1,n},c_{2,n},\dots,c_{M,n}]^\text{T}$ to maintain a slice. All slices are atomic and indivisible. Thus, the MNO manages its resources by adjusting the set of active slices, which can be presented as $\mathbf{s}=[s_1,s_2,\dots,s_N]^\text{T}$, where $s_n$ denotes the number of type-$n$ slices under maintenance. The allocation is subjected to the resource pool size:
	\begin{equation}\label{equ:res_feasibility}
		r_m-\sum_{n=1}^{N}c_{m,n}s_n\ge 0,\quad\forall m\in[1,2,\dots,M].
	\end{equation}
	We refer to the set of all allocations $\mathbf{s}$ that fulfill (\ref{equ:res_feasibility}) as the \emph{admissibility region} $\mathbb{S}$, which is obviously a finite set.
	
	\subsection{Tenant Requests and Slice Admission Control}\label{subsec:request_model}
	In SlaaS, slices are created and released upon requests from tenants. When a tenant requires a new slice to support its service, it sends a request to the MNO, which will be either accepted or declined by the MNO. Upon acceptance, the requested slice will be created and continuously maintained until the tenant requests to release it. Practically, both the requests for creation and for release arrive randomly. It is commonly assumed that:
	\begin{itemize}
		\item For every slice type $n$, the slice creation requests arrive as Poisson events with a rate of $\lambda_n$
		\item The lifetime of every type-$n$ slice is an independent $\mu_n$-mean exponential random variable.
	\end{itemize}

\begin{figure}[!htbp]
		\centering
		\includegraphics[width=.35\textwidth]{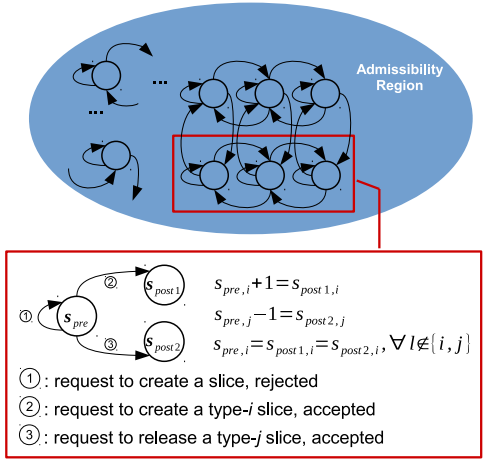}
		\caption{In asynchronous slice admission, the state is updated at request arrivals.}
		\label{fig:state_model}
	\end{figure}

	Thus, a state model can be proposed to describe the process of slice admission control, as illustrated in Fig. \ref{fig:state_model}. Each state represents a resource allocation $\mathbf{s}\in\mathbb{S}$, so that the MNO cannot make any decision that leads to an unfeasible resource allocation conflicting with (\ref{equ:res_feasibility}). Besides, note that requests for slice release are always accepted. 
	
	Generally, defining the incoming request $q$ and the MNO's binary decision $d$ as
	\begin{align}
		q&=\begin{cases}
			n&\text{request to create a type-$n$ slice}\\
			-n&\text{request to release a type-$n$ slice}
		\end{cases},\\
		d&=\begin{cases}
			1&\text{request accepted}\\
			0&\text{request declined}
		\end{cases},
	\end{align}
	respectively, the post-transition state $\mathbf{s}_\text{post}$ is a function of $q$, $d$ and the pre-transition state $\mathbf{s}_\text{pre}$:
	\begin{equation}
		\label{equ:async_state_transition}
		\begin{split}
			\mathbf{s}_\text{post}&=\mathcal{T}\left(\mathbf{s}_\text{pre},q,d\right)\\
			&=\left[s_{\text{pre},1},\dots,s_{\text{pre},\vert q\vert}+d\cdot\text{sgn}(q),\dots,s_{\text{pre},N}\right].
		\end{split}
	\end{equation}
	\subsection{Slicing Strategy}
	As aforementioned in Sec. \ref{subsec:request_model}, when receiving a tenant request $q$, the MNO makes a binary decision $d\in\{1, 0\}$. If $d$ is only a function of $q$ and the pre-transition network state $\mathbf{s}_\text{pre}$, we say that the MNO has a \emph{consistent slicing strategy} $d=D(q,\mathbf{s}_\text{pre})$. As the MNO cannot overload the resource pool or decline to release slices, given the admissibility region $\mathbb{S}$, a slicing strategy $D$ is valid only when
	\begin{align}
		&D\left(q,\mathbf{s}\right)=1,\quad&\forall q\in\{-1,-2,\dots,-N\}, \forall\mathbf{s}\in\mathbb{S}\label{equ:valid_strategy_1}\\
		&\mathcal{T}\left(\mathbf{s},q,D\left(q,\mathbf{s}\right)\right)\in\mathbb{S},\quad&\forall q\in\{1,2,\dots,N\}, \forall\mathbf{s}\in\mathbb{S}\label{equ:valid_strategy_2}
	\end{align}
	For any valid slicing strategy $D$, we have
	\begin{equation}
	\mathbf{s}_\text{post}=\mathcal{T}\left(\mathbf{s}_\text{pre},q,D\left(q,\mathbf{s}_\text{pre}\right)\right)=\mathcal{T}\left(\mathbf{s}_\text{pre},q\right).
	\end{equation}
	
\section{Synchronous Slice Admission Control Model}\label{sec:sync_model}

	\subsection{Synchronous Slice Admission}
	In the last section we have built an asynchronous state model of slice admission, where the updates of state, i.e. the responses of MNO, are triggered by requests arriving at random time. Practically, MNOs usually process requests in a framed approach, where all arrived requests, despite of the type, will be buffered in queue and sequentially responded at the end of every \textit{operations period}. This leads to a synchronous state model, where state updates are periodically triggered. The synchronous state model has self-evidently the same admissibility region $\mathbb{S}$ as the corresponding asynchronous state model, but complexer conditions for transitions between states.
	
	Normalizing the operations period length to $1$, the MNO makes sequential decisions to buffered requests at discrete time indexes. We use the vector $\mathbf{q}(t)=[q_1(t),q_2(t),\dots,q_{Q_t}(t)]$ to denote the request queue buffered during the $t^\text{th}$ operations period, where $Q_t$ is the total number of requests arrived in that period. Given an arbitrary strategy $d=D(q,\mathbf{s})$, the network state at $(t+1)$ is
	\begin{equation}
		\label{equ:sync_state_transition}
		\begin{split}
			\mathbf{s}(t+1)&=\mathcal{T}\left(\dots \mathcal{T}\left(\mathcal{T}\left(\mathbf{s}(t),q_1(t)\right),q_{2}(t)\right),\dots,q_{Q_t}(t)\right)\\
			&\overset{\Delta}{=}\tilde{\mathcal{T}}\left(\mathbf{s}(t),\mathbf{q}(t)\right),
		\end{split}
	\end{equation}
	where $\mathcal{T}(\cdot)$ is implemented according to (\ref{equ:async_state_transition}) w.r.t. $D(\cdot)$
	

	\subsection{Equivalent Markov Model}\label{subsec:eqvl_markov_model}
	The most important corollary of (\ref{equ:sync_state_transition}) is that: \emph{with memoryless distributions of request arrivals and a consistent slicing strategy, the synchronous state model of slice admission is Markovian}, as this subsection will show.
	
	As suggested in Sec.\ref{subsec:request_model} we assume that type-$n$ slice creation requests arrive as Poisson  events in rate of $\lambda_n$, and that the lifetime of every type-$n$ slice is an exponential random variable with mean of $\mu_n$. Thus, the probability mass function (PMF) that $k>0$ requests for type-$n$ slice creation  $q=n>0$ arrive during one operations period is
	\begin{equation}\label{equ:poisson_pmf}
		\text{Prob}(k_n)=\frac{\lambda_n^{k_n}e^{-\lambda_n}}{k_n!},\quad n\in[1,2,\dots,N].
	\end{equation}
	Given $s_n$ as the number of type-$n$ slices under maintenance, the PMF that $0<k_{-n}\le s_n$ requests for type-$n$ slice release $q=-n<0$ in the same operations period is
	\begin{equation}\label{equ:exponential_pmf}
		\text{Prob}(k_{-n}\ \vert\ s_n)=\frac{s_n!\left(1-e^{-\frac{1}{\mu_{n}}}\right)^{k_{-n}}}{k_{-n}!\left(s_n-k_{-n}\right)!e^{\frac{s_n-k_{-n}}{\mu_{n}}}}.
	\end{equation}
	We can merge them as
	\begin{equation}
		p_\text{A}(k_q,q\ \vert\ s_{\vert q\vert})=\begin{cases}
			\frac{\lambda_q^{k_q}e^{-\lambda_q}}{k_q!}&q>0\\
			\frac{s_{\vert q\vert}!\left(1-e^{-\frac{1}{\mu_{\vert q\vert}}}\right)^{k_{q}}}{k_{q}!\left(s_{\vert q\vert}-k_{q}\right)!e^{\frac{s_{\vert q\vert}-k_{q}}{\mu_{\vert q\vert}}}},&q<0
		\end{cases}
	\end{equation}	
	
	For convenience of reference, we define $\hat{\mathbf{q}}$ to denote the elements in a request sequence $\mathbf{q}$ \emph{regardless the order}. Because the arrival processes of different requests are independent from each other, the conditional probability that $\hat{\mathbf{q}}$ arrives during one operation period in the current network state $\mathbf{s}$ is
	\begin{equation}
		\text{Prob}\left(\hat{\mathbf{q}}\ \vert\ \mathbf{s}\right)=\prod\limits_{q\in\{\pm 1,\dots,\pm N\}}p_\text{A}\left(\sharp_{\hat{\mathbf{q}}}^q,q\ \vert\ s_{\vert q\vert}\right),
	\end{equation}
	where $\sharp_{\mathbf{x}}^x$ denotes the occurrence times of $x$ in $\mathbf{x}$. Note that every request arrival is independent from the others. Furthermore, the memoryless distributions guarantee that the arrival of every individual request remains consistent over the entire operations period. Thus, the arrival probability of a request sequence is obviously independent of its order (proven in \cite{chen1975poisson} as a feature of \emph{dependent trials}), i.e. 
	\begin{equation}\label{equ:order_independence}
		\text{Prob}(\mathbf{q}_1\ \vert\ \mathbf{s})=\text{Prob}(\mathbf{q}_2\ \vert\ \mathbf{s}),\quad \forall \{\mathbf{q}_1,\mathbf{q}_2\}: \hat{\mathbf{q}}_1=\hat{\mathbf{q}}_2.
	\end{equation}
	So we have
	\begin{equation}
			\text{Prob}(\hat{\mathbf{q}}\ \vert\ \mathbf{s})=\sum\limits_{i:\hat{\mathbf{q}}_i=\hat{\mathbf{q}}}\text{Prob}(\mathbf{q}_i\ \vert\ \mathbf{s})
			=Q!\times\text{Prob}(\mathbf{q}\ \vert\ \mathbf{s}),
	\end{equation}
	where $Q$ is the length of $\mathbf{q}$. This yields that
	\begin{equation}
	\text{Prob}(\mathbf{q}\ \vert\ \mathbf{s})	=\frac{\prod\limits_{q\in\{\pm 1,\dots,\pm N\}}p_\text{A}\left(\left.\sharp^q_\mathbf{\hat{q}},q\ \right\vert \ s_{\vert q\vert}\right)}{Q!}.
	\end{equation}	

	Now calling back (\ref{equ:sync_state_transition}), we are able to obtain the synchronous transition probability from state $\mathbf{s}(t)$ to $\mathbf{s}(t+1)$ as
	\begin{equation}\label{equ:sync_transition_prob}
		\text{Prob}\left(\mathbf{s}(t+1)\ \vert\ \mathbf{s}(t)\right)=\sum\limits_{\mathbf{q}: \tilde{\mathcal{T}}(\mathbf{s}(t),\mathbf{q})=\mathbf{s}(t+1)}	\text{Prob}(\mathbf{q}\ \vert\ \mathbf{s}(t)),
	\end{equation}
	which depends only on $\mathbf{s}(t)$. \emph{The synchronous slice management process is therefore Markovian}. As any other finite state Markov process, it can be characterized by an enumeration
	\begin{equation}
		f: \{1,2,\dots,\vert\mathbb{S}\vert\}\to\mathbb{S}
	\end{equation}
	and a corresponding transition probability matrix
	\begin{equation}
		\mathbf{P}=\begin{bmatrix}
		P_{1,1} & P_{1,2} & \dots & P_{1,\vert\mathbb{S}\vert}\\
		P_{2,1} & P_{2,2} & \dots & P_{2,\vert\mathbb{S}\vert}\\
		\vdots & \vdots & \ddots & \vdots\\
		P_{\vert\mathbb{S}\vert,1} & P_{\vert\mathbb{S}\vert,2} & \dots & P_{\vert\mathbb{S}\vert,\vert\mathbb{S}\vert}\\
		\end{bmatrix},
	\end{equation}
	where $P_{i,j}=\text{Prob}(f(j)\ \vert\ f(i))$ as defined by (\ref{equ:sync_transition_prob}).

	\subsection{Approximation in Practical Applications}\label{subsec:approx}
	There can be an infinite number of different request sequences $\mathbf{q}$ that fulfill $\tilde{\mathcal{T}}(\mathbf{s}(t+1),\mathbf{q})=\mathbf{s}(t+1)$, i.e. the calculation of $\mathbf{P}$, according to (\ref{equ:sync_transition_prob}), contains a traversal over an infinite domain of $\mathbf{q}$, which is computationally impossible.
	
	However, practically, if we set the operation period to a sufficiently short duration, after normalization the values of $\lambda_n$ will be small while $\mu_n$ being large for all $n$. In this case, the value of $p_\text{A}(k,q,\ \vert\ s_{\vert q\vert})$ fades out rapidly with increasing $k$ for all $q$, and $\text{Prob}({\mathbf{q}}\ \vert\ \mathbf{s})$ will therefore have non-negligible values \emph{only} for short request sequences $\mathbf{q}$. Thus, we can ignore all the cases with long sequences of arriving requests, and thereby limit the traversal operation in (\ref{equ:sync_transition_prob}) to a limited domain of $\mathbf{q}$, making it computationally feasible to solve $\mathbf{P}$.

	\subsection{Probability of Staying in a Specific State}
		Given a finite-state Markov chain which initiates from the state $\mathbf{s}(0)$ at $t=0$, with its transition probability matrix $\mathbf{P}$, the probability that it stays in a state $\mathbf{s}(T)$ after $T$ periods is
		\begin{equation}
			\text{Prob}(\mathbf{s}(T)\ \vert\ \mathbf{s}(0))=Q_{i,j}(T),
		\end{equation}
		where $i=f^{-1}(\mathbf{s}(0))$, $j=f^{-1}(\mathbf{s}(T))$ and
		\begin{equation}
			\begin{split}
				\mathbf{Q}(T)\overset{\Delta}{=}\mathbf{P}^T={\footnotesize\begin{bmatrix}
				Q_{1,1}(T) & Q_{1,2}(T) & \dots & Q_{1,\vert\mathbb{S}\vert}(T)\\
				Q_{2,1}(T) & Q_{2,2}(T) & \dots & Q_{2,\vert\mathbb{S}\vert}(T)\\
				\vdots & \vdots & \ddots & \vdots\\
				Q_{\vert\mathbb{S}\vert,1}(T) & Q_{\vert\mathbb{S}\vert,2}(T) & \dots & Q_{\vert\mathbb{S}\vert,\vert\mathbb{S}\vert}(T)\\
				\end{bmatrix}}.
			\end{split}
		\end{equation}

\section{Numerical Verification}\label{sec:verification}
\subsection{Environment Setup}
	To verify the feasibility of our proposed model, we setup a simplified slicing scenario with a normalized one-dimensional resource pool, i.e. $\mathbf{r}=[r_1]=[1]$. It supports to implement slices of an only type that each of them costs a resource of $c_{1,1}=0.3$. Under this specification, only $2$ types of request and $4$ states are available:
	\begin{align}
		&q=\pm 1,\\
		&\mathbb{S}=\{\mathbf{s}_1,\dots\mathbf{s}_{4}\},\quad\mathbf{s}_i=[i-1].
	\end{align}
	There are in total $2^{2\times4}=256$ different constructions of $D$ under such specifications. Nevertheless, according to (\ref{equ:valid_strategy_1}) and (\ref{equ:valid_strategy_2}), only $2^3=8$ out of them are valid slicing strategies. To study the relation between the precision of our model and the request arrival rate, we designed three different slice service demand scenarios, as listed in Tab. \ref{tab:scenarios}.
	\begin{table}[!htpb]
		\centering
		\begin{tabular}{c|c|c|c}
			\toprule[1.5px]
			\textbf{Scenario}&A&B&C\\\hline
			\textbf{Request arriving rate for new slices} ($\lambda_1$)&1&0.8&0.5\\\hline
			\textbf{Average slice lifetime }($\mu_1$)&4&4&4\\
			\bottomrule[1.5px]
		\end{tabular}
		\caption{Three different service demand scenarios}
		\label{tab:scenarios}
	\end{table}

\subsection{Simulation Procedure and Results}
\begin{figure}[!htbp]
		\centering
		\includegraphics[width=.43\textwidth]{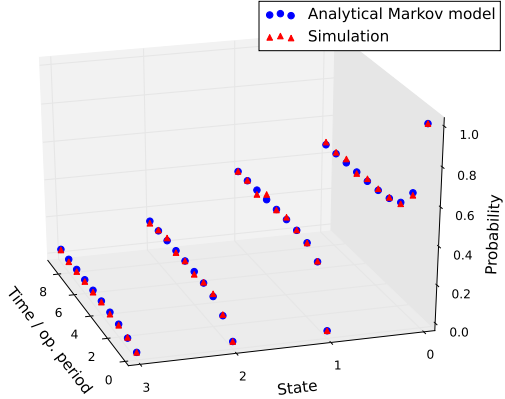}
		\caption{The estimated and simulated PMFs of network state in the first $10$ operation periods. The resource pool was initialized fully idle, an ``always accept'' strategy was taken in the reference scenario C, $Q_{\max{}}^+=4$ was considered.}
		\label{fig:case_study}
	\end{figure}

	Given a certain scenario and a certain strategy, the state transition probability matrix $\mathbf{P}$ can be estimated as described in Sec. \ref{subsec:eqvl_markov_model}. As discussed in Sec. \ref{subsec:approx}, we must set an upper-bound $Q^+_{\max{}}$ to the possible arrival number of slice creation requests, so that $\mathbf{P}$ could be approximated by an estimation $\hat{\mathbf{P}}$ with a limited effort of computation. Fig.~\ref{fig:case_study} shows an example result in the first ten operations periods with fixed initial state, service demand scenario and decision strategy. An excellent match between the analytical estimation based on our proposed Markov model and the simulation results, verifying the feasibility of our proposed approximate estimator $\hat{\mathbf{P}}$.
	
\begin{figure}[!htbp]
		\centering
		\includegraphics[width=.43\textwidth]{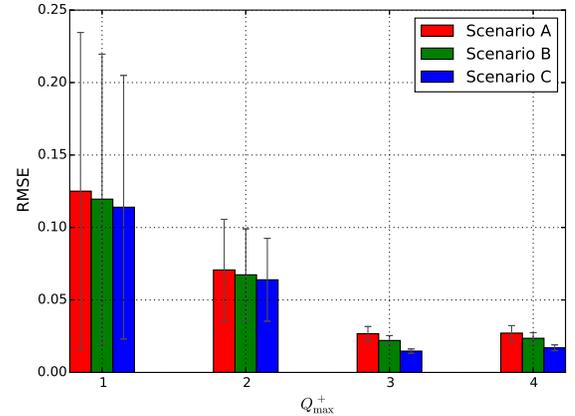}
		\caption{The estimation error of state transition probability matrix $P$ with respect to the $Q_{\max{}}^+$ configuration, the strategy and the service demand scenario.}
		\label{fig:sim_result}
	\end{figure}

	Then, to evaluate the estimation accuracy, we obtained the empirical transition probability matrix through simulations. For every specification set of scenario, strategy and $Q^+_{\max{}}$, 1000 independent runs of Monte-Carlo test were carried out. In each individual run, the network was first initialized to a random state, then operated with the specified slicing strategy for 100 operations periods. We recorded the system state in every operation period, and thereby obtained the empirical transition probability matrix $\tilde{\mathbf{P}}$. The root of mean square error (RMSE) of the estimation $\hat{\mathbf{P}}$ is thus evaluated as:
	\begin{equation}
	\begin{split}
	\epsilon=\sqrt{\frac{1}{\vert\mathbb{S}\vert^2}\sum\limits_{i=1}^{\vert\mathbb{S}\vert}\sum\limits_{j=1}^{\vert\mathbb{S}\vert}\left[\frac{2(\hat{P}_{i,j}-\tilde{P_{i,j}})}{\hat{P}_{i,j}+\tilde{P_{i,j}}}\right]^2}
	\end{split}
	\end{equation}
	Note that the value of $\epsilon$ depends on the decision strategy $D(\cdot)$, the preset upper-bound $Q_{\max{}}^+$ and the request statistics. So we repeated the simulation 4 times with $1\le Q_{\max{}}^+\le 4$, respectively. For every value of $Q_{\max{}}^+$, we evaluated $\epsilon$ for all 8 valid strategies in all 3 aforementioned reference service demand scenarios.
	The results are depicted in Fig.~\ref{fig:sim_result}. It can be seen that independent of the scenario, both the mean and the variance of estimation error sink quickly to a negligible level as $Q_{\max{}}^+$ increases, which significantly supports our analysis on the Markov model in Sec.~\ref{sec:sync_model}.
	We can also observe a dependency of the RMSE on the scenarios, that the error decreases along with the arrival rate of requests for new slices, which can be easily explained by the point that a higher request arrival rate leads to higher probability of long request sequences, which indicates a degrade of approximation accuracy with the same configuration of $Q_{\max{}}^+$. \emph{This also suggests to select a shorter operations period in practical application, in order to limit the required computational effort.}

\section{Conclusion and Discussions}\label{sec:conclusion}
In this paper, we have proposed a state model of synchronous slice admission in 5G mobile communications networks. We have proved that when the statistics of request arrivals are memoryless, e.g. when they are Poisson / exponential processes, and when the MNO takes a consistent slicing strategy, the state model is Markovian. An approximate analytical calculation of the state transition matrix is then proposed and verified by numerical simulations.

We would like to highlight one outcome of this work that under certain service demand statistics, the Markov model of the network state is individually determined by the applied slicing strategy. This guarantees that the optimization problems of slice admission can be equivalently transformed into the problem of searching the best slicing strategy.

It is also worth to note that the space of valid slicing strategy can dramatically grow to a huge size as the slice pool and number of slice types increase, and thus calls for fast and efficient optimizing algorithms, which deserves further study in the future.

\bibliographystyle{IEEEtran}
\bibliography{references}

%
%

\end{document}